\documentclass[%
reprint,
superscriptaddress,
nofootinbib,
amsmath,amssymb,
aps,
pre,
]{revtex4-2}

\usepackage{gensymb,units,textcomp,color,leftidx,xspace,epsfig,units,bm,graphicx,mathrsfs}
\usepackage{hyperref}
\usepackage[acronym]{glossaries}
\usepackage[caption=false]{subfig}

\newcommand{\UVa}{Departamento de F\'isica Te\'orica At\'omica y \'Optica, Universidad de Valladolid, 47011 Valladolid, Spain}
\newcommand{\ULPGC}{iUNAT-Departamento de F\'isica, Universidad de Las Palmas de Gran Canaria, 35017 Las Palmas de Gran Canaria, Spain.}

\newacronym{FMD}{FMD}{Full Molecular Dynamics}
\newacronym{TMD}{TMD}{Trivial Molecular Dynamics}
\newcommand{\FMD}{\gls{FMD}\xspace}
\newcommand{\TMD}{\gls{TMD}\xspace}
\newcommand{\rmin}{a}

\begin{document}

\title{Tracking electron capture process in classical molecular dynamics simulations for spectral line broadening in plasmas}
\author{D. Gonz\'alez-Herrero}
    \affiliation{\UVa}
\author{G. P\'erez-Callejo}
    \affiliation{\UVa}
\author{R. Florido}
    \affiliation{\ULPGC}
\author{M. A. Gigosos}
    \email{gigosos@coyanza.opt.cie.uva.es}
    \affiliation{\UVa}
    
\date{\today}

\begin{abstract}
	Plasma spectroscopy is a fundamental tool for diagnosing laboratory and astrophysical plasmas. Accurate interpretation of spectra depends upon precise modeling and comprehension of Stark-broadening and other mechanisms affecting spectral lines. In this context, computer simulations have emerged as valuable tools, offering \textit{idealized experiments} with well-defined conditions. Molecular dynamics simulations, in particular, excel at replicating the particle interactions within the plasma and their impact on the state of a radiating atom or ion. However, these simulations present challenges in tracking electron capture processes, since setting an unambiguous criterion to distinguish between bound and free electrons is not trivial. In this paper we introduce a new algorithm that, within a classical framework, precisely identifies the scenario in which an electron is captured by an ion and then follows a stable orbit around it. The algorithm's applicability extends to emitters with charges $Z\ge 1$. The procedure enables the correct identification of valid time-histories of the electric microfield perturbing the emitting ion, which will be used for subsequent line shape calculations. The ionization balance results obtained from the application of this algorithm are compared with an additional method based on the potential energy of the particles in the simulations and with atomic kinetic simulations. For both methods, we find good agreement, therefore validating the use of this approach.
\end{abstract}

\maketitle

\section{Introduction}
\label{sec:Intro}

Accurate modeling of spectral line shapes is indispensable for investigating the properties of high energy density plasmas. The analysis of emission or absorption spectra, which typically involves the comparison of experimental data with spectroscopic models, stands as a key diagnosis tool for laboratory and astrophysical plasmas.  In particular, spectral line shapes, broadened and shifted by the kinetics of plasma particles and their interactions, provide relevant information about the plasma temperature and density. Spectral lines arise from atomic transitions between bound energy states of a radiating ion or neutral atom, the \textit{radiator}. 

In the context of hot and dense plasmas the Stark effect emerges as the dominant broadening mechanism \cite{griem1974}. Unlike the isolated scenario, when embedded in a plasma, the radiator is subject to the electric field generated by its surrounding charged particles during the time interval in which it emits. The shape of the spectral lines in this case depends on the plasma-induced electric field, its time evolution (tied to the plasma particle dynamics) and the radiators' ensemble.

Despite significant strides in recent decades \cite{gigosos2014,gomez2022}, modeling spectral line shapes remains a formidable challenge owing to its interdisciplinary nature, requiring the integration of various ingredients from atomic physics, plasma physics, statistical mechanics, collision theory, and computational physics. Moreover, in the high-density regime, dedicated validating experiments are notably scarce, presenting a substantial obstacle to grounding theoretical models in empirical reality. 

Generally, two primary approaches are used for line-broadening calculations: analytic methods and computer simulations. Analytic models navigate the complexities of line broadening by incorporating assumptions and mathematical manipulations to handle the stochastic nature of line emission. One common approximation involves the separate calculation of electron and ion broadening contributions. Since ions are much heavier than electrons, they are assumed to be static over the relevant electron time scales. The analytic approach then puts the effort in obtaining an accurate representation of the electron broadening operator. In this regard, some common approximations include the simple and well-known \textit{impact} approximation, or the \textit{second-order} approximation as a simplified way to compute the electron broadening operator \cite{griem1974, gigosos2014}. Once the electron broadening contribution is computed for a given ion distribution, i.e. for a given value of the ion microfield, the line emission is integrated over the statistical distribution of the ion microfield for the conditions of interest, a facet that entails its own extensive research area in microfield probability distributions. The inherent advantages of analytic methods lie in their computational efficiency, as the time required to calculate a line shape is generally much shorter than for computer simulations, and they operate without numerical noise. 

On the other hand, computer simulations excel at replicating plasma particle interactions and their impact on the state of a radiating atom or ion \cite{stambulchik2010}, proving particularly advantageous in gaining insights into the line broadening problem. In this case, simulated plasma particles traverse a confined space, following classical trajectories in discrete timesteps. These classical particles generate a time-dependent Coulomb or Debye potential or, equivalently, a time-sequence of the electric field. The emitting atom or ion then internally evolves subject to the effect of this electric field. This evolution is used to numerically solve the time-dependent Schr\"odinger equation to obtain the corresponding atomic spectrum.

While computer simulations have become a standard tool to study Stark-broadened lineshapes, there are some physical broadening mechanisms that are still unaccounted for in these type of calculations. Among them is the so-called \textit{recombination broadening}, produced when a free electron within the plasma becomes bound to an emitting ion, therefore stopping its emission. Although this effect can be accounted for in analytic methods \cite{gomez2020}, including it in computer simulations remains a challenge. 

It is in this context that our work is framed. We present an algorithm to accurately track the electron capture processes within computer simulations, key to obtaining correct time-sequences of the electric field within the plasma and account for the additional recombination broadening. 

This manuscript is structured as follows: in Section \ref{sec:ComSim}, we describe computer simulations and give an overview of the general operation of \FMD simulations for Stark broadening. Then, in Section \ref{sec:Criterion} we present the criterion for detecting electron captures within \FMD simulations. An example of the application of this criterion to a computer simulation is presented in Sec. \ref{sec:Example}, and the results are compared and validated with additional metrics in Sec. \ref{sec:ionization}. Finally, Sec. \ref{sec:Conclusions} summarizes the conclusions and limitations of this work and explores future applications of this type of techniques.

\section{Overview of computer simulations}
\label{sec:ComSim}

Computer simulations for calculating Stark-affected spectra started with the work of Stamm \cite{stamm1979}. They usually follow these steps \citep{gigosos2014}:
\begin{itemize}
	\item [a)] In a computer, one reproduces a plasma composed of charged and/or neutral particles which are distributed in a given region of space. These particles move according to certain pre-established statistical distributions which arise from a given physical configuration (electron density and temperature). The simulation evolves in discrete timesteps and the calculation permits the description of the position and velocity of the particles at every timestep.

    \item[b)] The simulated plasma includes the emitters that are considered for the spectral calculations. For each timestep, the electric field produced by all plasma particles on a given emitter is calculated. This is the time-dependent field sequence that this work focuses on.
    \item[c)] The emitting atom or ion internally evolves, subjected to the effect of this electric field. By numerically solving the time-dependent Schr\"odinger equation, one obtains the corresponding evolution operator $U(t)$. It is then possible to calculate the dipolar moment of said emitter as a function of time, $D(t) = U^\dagger(t)D(0)U(t)$. One then obtains the autocorrelation function of this dipolar moment as $C(t)=\text{Tr}\left[\rho D(t)D(0)\right]$, where $\rho$ is the density matrix (usually taken as the identity matrix in Stark broadening calculations).  
    \item[d)] This process is repeated a sufficient number of times, to average over a sufficiently large ensemble of field sequences. The spectrum of interest is then obtained as the Fourier transform of the average autocorrelation function of the dipolar moment, $\left\{C(t)\right\}$, i.e.
	\begin{equation}
		\label{eqn:spectrum}
	  I(\omega)\propto \text{Re}\int_0^{\infty}\left\{C(t)\right\}e^{i\omega t} dt
	\end{equation}
\end{itemize}

Simulations often simplify the problem and reduce computational time by employing straight-line path trajectories for particle movement, i.e. with the simulated particles being independent. We will refer to this type of simulations as \TMD. 
\TMD calculations are useful when the typical electric-interaction energies are much lower than the mean kinetic energy of particles (a weakly-coupled high temperature plasma) and, specially for non-charged emitters \cite{stamm1979, gigosos1987, hegerfeldt1988, rosato2009, gigosos2014SIMULA, gomez2016}. 

Noteworthy exceptions to this norm are \FMD simulations, in which the calculation takes into account the electric interaction of all particles in the plasma. This constitutes a more accurate (and more computationally expensive) method, which must solve Newton's equations of motion to determine the trajectories of all particles.  \FMD simulations are necessary for medium and strongly-coupled plasmas and, specially when the emitter is an ion, given that its interaction with the surrounding particles in the plasma affects both its own trajectory and those of the other particles, thus modifying the time-evolution of the electric field of interest \cite{ferri2007,stambulchik2007,gigosos2021} \footnote{It is worth mentioning that there is an intermediate type of calculations, in which the perturbers are treated as independent particles, but their interaction with the emitter is taken into account explicitly. This results in the charged perturbers moving in hyperbolic trajectories when a Coulomb field is considered, or in slightly different paths if Debye screening is introduced, in which case the corresponding differential equation must be solved numerically \cite{stambulchik2006}. }.

While it is clear that \TMD models cannot reproduce recombination broadening, as the particles are considered independent, it is possible to use \FMD simulations to detect when an electron becomes bound to an ion and account for this additional broadening.

\section{Criterion for detecting trapped electrons in a computer simulation}
\label{sec:Criterion}

As discussed above, in \FMD simulations, the particle trajectories are solved by accounting for the Coulombian interaction between all particles in the plasma. 
As the simulation evolves, it becomes possible that the binding energy (negative) between an electron and an ion becomes too large - in absolute value - with respect to their kinetic energies (positive), so that the electron becomes \textit{trapped} by the ion. 
In that case the electron becomes bound and its trajectory follows an orbit around the ion.
The two particles in this pair can therefore no longer be considered free and must be labeled as a neutral atom, or as an ion with a positive charge lower than that of the free ion. 

This circumstance was extensively discussed in our previous work \cite{gigosos2018}, where we developed a statistical model for the equilibrium of the simulated plasma, taking into account the process of electron capture and release by the ions (recombination and ionization respectively). 
The model developed in that work, perfectly reproduced what is observed in the simulation, thus permitting the correct identification of the number of free electrons in the plasma, considering that these \textit{trapped} electrons do not contribute to its value.
For this model, it was necessary to establish a clear criterion to determine when an electron-ion pair must be considered bound, and when those particles must be considered free. 
While in \citet{gigosos2018} ions of charge $+e$ were used (where $e$ is the electron charge), in this work we are interested in plasmas with charge $+Ze$, with $Z>1$. 
The purpose of this work is then to extend this criterion to situations in which the number of trapped electrons can exceed $1$.

The need for such a criterion is obvious. In the simulations, it is necessary that the density of free electrons is a given value (generally predetermined). To calculate such value, one needs to \textit{count} the electrons that are really free in the simulated plasma. This number does not match that of the electrons included in the calculations, since some of them will become bound to the ions.

Furthermore, there is another important reason that requires knowing when an ion has captured an electron and when it can be considered free. Let us consider a \FMD simulation in which we are interested in calculating the spectrum from an ionic emitter with charge $+Ze$. To that end, during the simulation process, we must register the temporal sequence of the electric field \textit{seen} by that charged emitter, which is produced by the ions and free electrons surrounding it. If, at a given moment, an electron is captured by the ion of consideration, from that moment on, the electron will orbit around it at a short distance. Therefore, the electric field felt by that emitter will be an oscillating function with a large amplitude (since the oscillating electron is very close), which is non-physical and thus spoils the field sequence used for the calculation of the spectrum. 

In addition, the capturing ion, originally with charge $+Ze$ becomes now, along with the trapped electron, an ion of charge $+(Z-1)e$ and can therefore no longer be considered a contributor to the spectral line in which we are interested. 
In the framework of semiclassical computer simulations it is then necessary to wait for the corresponding ion to recover its original charge in order to continue using it for the calculation.
The moment the electron is captured ($t_{cap}$) determines when the ion changes its \textit{species} and therefore the emission coherence is abruptly lost, making the correlation function of the dipole moment fall to zero ($C(t>t_{cap})=0$). 
This causes a broadening of the line by the recombination process, which highlights one of the advantages of these calculations \citep{gigosos2021}. 
As a consequence, in practice, the corresponding time-sequence of the electric field is \textit{cut} in the simulation process. 
Since the duration of the different field sequences varies and, as a consequence the duration of the dipole autocorrelation function varies, in order to perform the average of the autocorrelation functions, indicated in Eqn. \ref{eqn:spectrum}, these need to be zero-padded to the same length.

Usually, electrons are labeled as \textit{bound} to an ion when their distance is lower than a given value $a$ (whose value is discussed later) and the total energy (kinetic plus potential) of the electron-ion pair is negative. While this criterion \citep{natitesis} works well in most cases, it presents certain problems that become more important for ions with charge $Z>1$ and more strongly coupled plasmas.

For example, if one follows this criterion, it is possible for an electron to orbit at distances larger than $\rmin$ (and thus be considered \textit{free}), but still produce a non-physical oscillating electric field on the ion.  
On the other hand, orbiting electrons can spend part of their orbit closer than $\rmin$, and part of it at larger distances. 
In these cases, ions with an orbiting electron would be considered free during part of the electron orbit, and bound during the rest. This woul produce unphysically short sequences of the electric field, overestimating the recombination broadening.

Furthermore, for ions with charge $Z>1$,
which can have more than one bound electron, this criterion becomes impractical  \citep{diegotesis}. 
Given that it considers only the instantaneous conditions of the simulation, if two electrons were orbiting an ion, this method would only detect them if the two of them were closer than a distance $\rmin$ at the same time (which becomes more unlikely as $Z$ increases). 

It is then clear that, to eliminate the non-physical field sequences produced by oscillating electrons and correctly compute the number of bound electrons (necessary to correctly obtain the free electron density), it is necessary to establish a criterion for the \FMD simulations that considers, not only the instantaneous characteristics of the electron-ion pairs, but also the associated time-history.

To correctly select the valid field sequences we choose a global criterion, in which whether an ion has trapped an electron or not, not only depends on its conditions at a given time, but also on its history. For a given ion, we store (for each step of the simulation) its potential energy, the electric field at its position and the \textit{label} (an index identifying each electron in the simulation) of the $Z+1$ closest electrons, where $Ze$ is the charge of the emitter. 

In order to correctly calculate the potential energy, since the interaction between particles is Coulombian, it is necessary to apply a \textit{regularization} to the ion-electron interaction potential to avoid divergencies at short distances. While several procedures for this regularization have been reported (see \citet{bonitz2023} and references therein), in this work we will use that described in \citet{gigosos2018}, which considers a parabolic potential for interparticle distances below a given value $\rmin$ and hyperbolic for larger distances. This method naturally accounts for the diffraction of electrons around an ion and determines an effective atom size. We used the following expression for $Z\geq1$ ions, rescaled from \citet{gigosos2018}

\begin{equation}
\label{eqn:Potencial}
    V_{ie}(r) = 
        \begin{cases}
		V_b\left[\frac{1}{3}\left(\frac{r}{a}\right)^2-1\right],    &\quad r\leq a\\
            -\frac{Ze^2}{4\pi\varepsilon_0 r},   &\quad a<r\leq R_I\\
            0,  &\quad r>R_I
        \end{cases}
\end{equation}
where $V_b$ is the energy of the bottom of the potential well and $R_I$ is the radius of the interaction sphere, defined as half the size of the cubic simulation box. For ions with charge $Z$, $\rmin$ is a parameter given by
\begin{equation}
	\rmin = \frac{3}{2}\cdot\frac{Ze^2}{4\pi\varepsilon_0 V_b}.
\end{equation}

The parameter $\rmin$ defines the distance at which the potential changes from Coulombian to parabollic. Its physical meaning, as stated in Reference \cite{gigosos2018}, ``\textit{corresponds to the case of having the ion charge uniformly distributed in the volume of a sphere with radius $\rmin$, which is also permeable to pointlike electrons.}'' 
The relation between this size and the available volume per electron in the plasma, is the characteristic parameter for the equilibrium constant appearing in the Saha equation (see Equation 23 in Reference \cite{gigosos2018}).

In order to determine then the value of $V_b$, we impose the condition $V_{ie}(\rmin) = -V_i$, where $V_i$ is the energy that would be required to ionize one electron that becomes bound, i.e. the ionization energy of the $Z-1$ charge state.
In Section \ref{sec:ionization}, we will discuss how this condition yields results for the classical simulation which are consistent with those obtained with the quantum Saha equation. By doing this, one obtains that

\begin{equation}
	V_b = \frac{3}{2}V_i,
	\label{eqn:V0}
\end{equation}
and
\begin{equation}
	\rmin = \frac{Ze^2}{4\pi\varepsilon_0 V_i}.
\end{equation}

The first step to establish a criterion for when an ion has trapped an electron is then to track down the moments when the potential energy of the emitter, $\mathscr{E}_p$, is lower than a given threshold energy $\mathscr{E}_{th}$.
The energy $\mathscr{E}_p$ here is the \textit{full potential energy} of the emitter, that is, calculated by considering the interaction with all other particles in the simulation, taking into account the regularization of the ion-electron interaction potential.
In particular, to keep consistency with the original criterion for $Z=1$ atoms, we choose
\begin{equation}
    \mathscr{E}_{th} = - V_i,
    \label{eqn:ETh}
\end{equation}
which corresponds to the potential energy of an ion with an electron at a distance $\rmin$.

Whenever the ion potential energy is lower than $\mathscr{E}_{th}$, the electron that is closest to it (its first neighbor) is identified. Then, the time interval during which that electron has been among the $Z+1$ closest neighbors is considered. If this interval is longer than a given threshold value $\tau_{\text{bound}}$, then the average total energy of the electron-ion pair in that interval is calculated. If the pair has a negative total energy during that interval, then the electron is considered to be bound for the whole interval. If not, it is considered that the negative energy seen by the emitter is caused by a strong collision, since the electron causing it did not remain around the ion. It is worth noting that the $Z+1$ closest neighbors are considered, since it is possible that a fast electron gets closer to the ion core than other bound electrons for a short period of time, without freeing them.

Since the full time-history of the ion is needed in order to check whether an electron is bound or free, its history must be saved so that it can be accessed when the energy criterion is met, and the correct field sequences can be extracted.

\subsection*{Determining the timescale for bound electrons}

In the classical picture of these \FMD simulations, an electron orbiting an ion will do so with a given period, such that the orbit is stable. For a hyperbolic potential (such as the Coulomb potential), the condition for a stable circular orbit is given by
\begin{equation}
    m_e v^2 = \frac{Ze^2}{4\pi\varepsilon_0 r},
\end{equation}
where $m_e$ is the electron mass, $v$ its velocity, and $r$ the orbit radius. Under this condition, an electron orbiting an ion at the typical thermal velocity ($v_{T} = \sqrt{2k_BT_e / m_e}$, with $k_B$ the Boltzmann constant and $T_e$ the electron temperature) will do so at a distance of
\begin{equation}
    r_{T} = \frac{Ze^2}{8\pi\varepsilon_0 k_B T_e},
\end{equation}
and therefore, the corresponding orbital period is given by
\begin{equation}
    \tau_{T} = \frac{2\pi r_{T}}{v_{T}} =  \frac{Z e^2}{2\varepsilon_0}\cdot\frac{m_e^{1/2}}{(2k_B T_e)^{3/2}}.
\end{equation}

This expression can be simplified by defining the typical timescale of electron dynamics $t_0$ as the time it takes an electron moving with the characteristic thermal velocity $v_T$ to travel the typical distance between electrons ($r_e = (4\pi n_e/3)^{-1/3}$). Using this definition, one gets
\begin{equation}
	t_0 = \frac{r_e}{v_T} = \left(\frac{3}{4\pi n_e}\right)^{1/3}\cdot\left(\frac{m_e}{2k_BT_e}\right)^{1/2},
\end{equation}
which yields
\begin{equation}
	\tau_T = \frac{\pi}{3} Z\rho^2 t_0,
\end{equation}
where $\rho$ is the coupling degree of the plasma, given by the relation between the typical distance between electrons and the Debye radius.

We take this characteristic orbit time as a basis to consider whether an electron is bound to an ion. 
Although it would be possible to consider $\tau_{T}$ as the temporal threshold -- that is, if an electron is within the $Z+1$ closest neighbors of a given ion for at least $\tau_{T}$, and its total energy is negative, it is considered bound -- this is not necessary, since an electron that loops around an ion once, although it can be technically trapped, spends too short a time orbiting to break the correlation of the atomic dipole completely, and does not create an oscillating electric field.
In practice, these events behave equivalently to strong collisions within the plasma.

To account for this possibility, we take $\tau_{\text{bound}}=3\tau_{T}$ as the threshold for considering an electron as bound to a given ion.
We find that this value produces good results that are self-consistent within the simulation (as we will discuss in Section \ref{sec:ionization}).

We note here that the above expression for $\tau_{T}$ was obtained for a hyperbolic potential. However, as indicated in Eqn. \ref{eqn:Potencial}, in our simulations, the potential becomes parabolic for distances $r<\rmin$, and thus, for this expression to be valid, it is necessary that $r_{T} > \rmin$. By comparing the above expressions it can easily be seen that this condition is equivalent to
\begin{equation}
	V_i > 2 k_B T_e.
\end{equation}
However, this is always the case when recombination effects are relevant given that, in reality, if the temperature is high enough so that it becomes comparable with the ionization potential, the electrons in the simulation are all free, and recombination effects are negligible (as we will show in the following section). 

\begin{figure*}
    \centering
    \includegraphics[width=\linewidth]{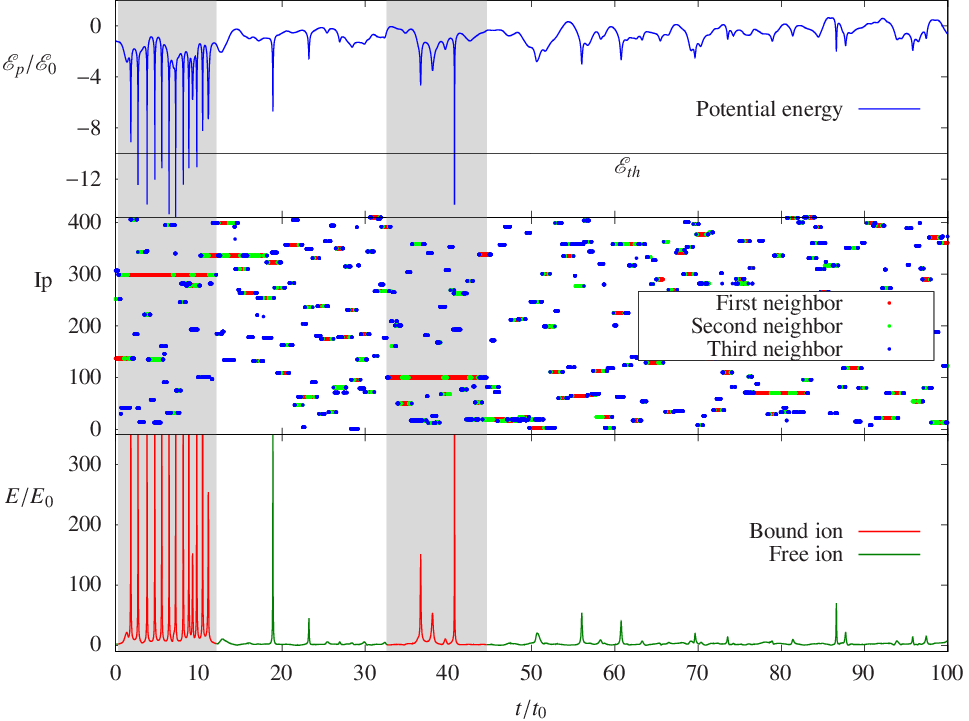}
    \caption{Time-history of a plasma with $\rho = 0.6$ and emitters with charge 2. From top to bottom: potential energy of an emitter as a function of time, indices of the three closest electrons at each moment, and electric field. The shaded regions correspond to the moments when a particle is trapped. The potential energy is given in units of $\mathscr{E}_0$, a characteristic energy of the simulation, and the electric field in units of $E_0$, the electric field produced by a single electron at the typical distance $r_e$.
    }
    \label{fig:E1}
\end{figure*}

\section{Results and discussion}

\subsection{Application of the criterion}
\label{sec:Example}

Let us now show an example of the application of this method, for the particular case of ions with charge $Z=2$, which means that the three closest neighbors must be monitored. In this example, the coupling parameter of the plasma is $\rho = 0.6$, and therefore $\tau_{T} \sim 0.75 t_0$ and $\tau_\text{bound} \sim2.25t_0$. Figure \ref{fig:E1} shows the time-history of the potential energy of an emitter (top), the indices $I_p$ of the three electrons that are closest to it at every timestep in the simulation (middle) and the value of the modulus of the electric field acting on said emitter (bottom panel). The potential energy is given in units of the characteristic energy of the simulation $\mathscr{E}_0$, a parameter given by the average kinetic energy of the particles at the beginning of the simulation --before equilibrium is reached--, as further detailed in \citet{gigosos2018}. The electric field is given in units of $E_0$, i.e. the field produced by a single electron at the typical distance $r_e$.

At $t\sim2.7t_0$, when the potential energy becomes lower than $\mathscr{E}_{th}$, the electron that is closest to the emitting ion is electron number 298. We then observe that this electron has been among the three closest neighbors from $t\sim 0$ to $t\sim 10t_0$, approximately 10 times the characteristic time of the simulation. This is much longer than the typical duration of a collision, so we must assume that this corresponds to a bound electron. Indeed, if we now consider the electric field between $t\sim0$ and $t\sim10t_0$, it oscillates periodically, with a period of approximately $0.75t_0$ (which is the value of $\tau_{T}$ in this case), reaching very high values. A similar situation is observed at $t\sim40t_0$, although in this case the trapped electron oscillates around the ion only a few times, and therefore its imprint on the electric field history is not as clear.

It can be seen now that strong collisions and electron captures appear in the time-history of the potential energy of the ion as clear spikes, and what determines whether an ion has captured an electron or not, is the time-dependent part of the criterion. For this reason, the value of $\mathscr{E}_{th}$ must be one such that every electron capture event is detected.
In the example presented in Fig. \ref{fig:E1}, the ionization potential is $V_i = 10\mathscr{E}_0$ (corresponding to a potential well of $V_b = 15\mathscr{E}_0$), which following the definition in Eqn. \ref{eqn:ETh} corresponds to $\mathscr{E}_{th} = -10\mathscr{E}_0$. However, if one were to use a threshold energy with a lower absolute value, the same results would have been obtained, although the condition $\mathscr{E}_p<\mathscr{E}_{th}$ would have been fulfilled more times, and therefore, the time-history would have had to be checked more often. 
In fact, by taking a lower threshold energy, the only difference is the computational cost of the simulation. For example, at $t\sim20t_0$, the potential energy jumps below $6\mathscr{E}_{th}$, as an electron passes very close to the emitting ion. If the chosen threshold energy were lower, the algorithm would inspect this event as a potential electron capture. However, the colliding electron in this case does not remain among the three closest neighbors for long enough to be a bound particle and, indeed, the electric field history confirms that this corresponds to an isolated collision.
It can therefore be seen that by applying this criterion, it is possible to identify and select the appropriate sequences of the electric field that a radiator is subjected to during its emission, including the correlation loss owed to capturing an electron. These sequences can then be further used for Stark-broadening line shape calculations. 

It is worth now to examine the difference between this new criterion and the original criterion for bound electrons \cite{natitesis}. Originally, the time-history of the simulation was not considered, and the field sequences were cut every time a bound electron was detected, starting a new field sequence immediately afterwards. Examining Fig. \ref{fig:E1}, we can see that, using this criterion would result in the generation of $\sim 10$ different field sequences between $t\sim 0$ and $t\sim 10t_0$, each $\sim 0.75t_0$ in duration (every time the oscillating electron gets sufficiently close to the ion). These short sequences are unphysical, since they are produced by an oscillating semiclassical electron and would result in an overestimation of the recombination broadening. By considering the time-history of the system, with the new criterion proposed here, these unphysical field sequences are removed.

\begin{figure*}
  \begin{center}
    \includegraphics[width=\columnwidth]{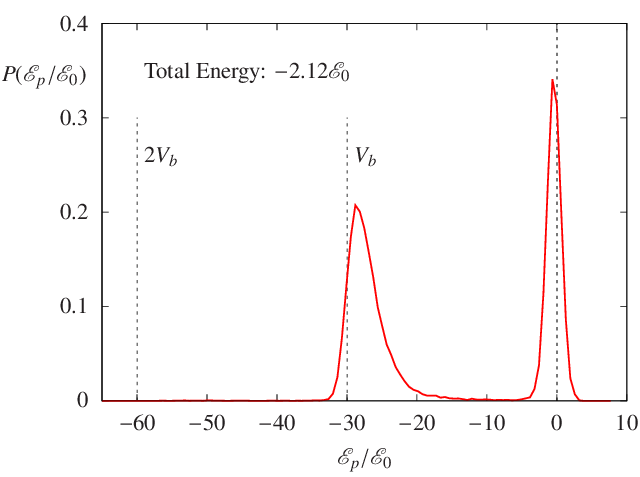}
    \hfill
    \includegraphics[width=\columnwidth]{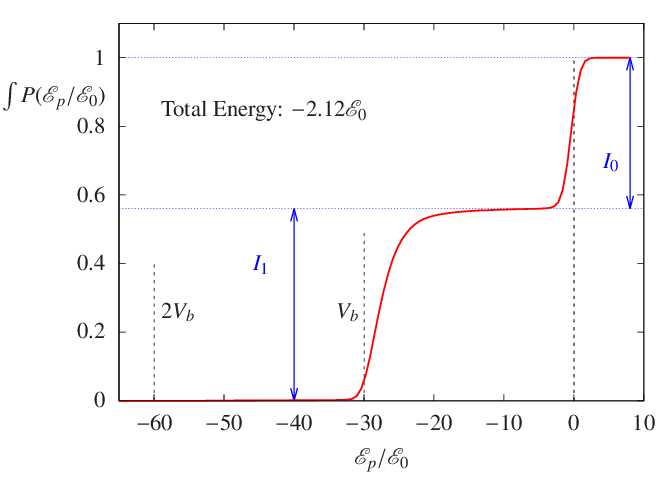}
	  \caption{Distribution of the potential energy of ions (left) and its integral (right). Data correspond to a Helium plasma ($Z=2$), with $V_b = \unit[30]{\mathscr{E}_0}$, and a coupling parameter $\rho_N = 1.0$. The mean total energy per particle of the system is $\mathscr{E}_t = -2.12\mathscr{E}_0$.}
    \label{fig:418}
  \end{center}
\end{figure*}

\begin{figure*}
  \begin{center}
    \includegraphics[width=\columnwidth]{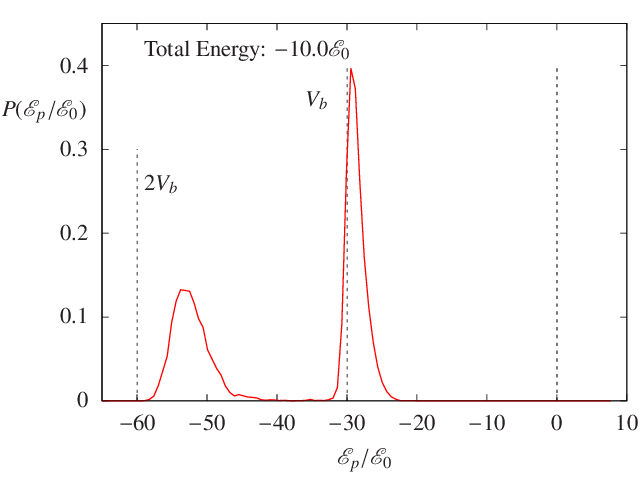}
    \hfill
    \includegraphics[width=\columnwidth]{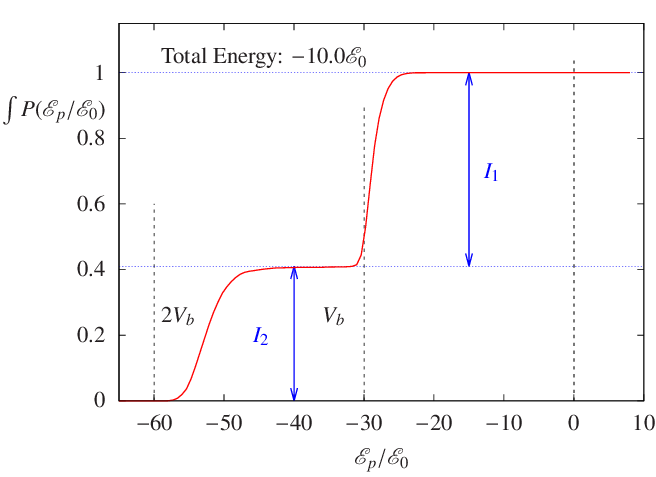}
    \caption{Same as in Fig. \ref{fig:418}, with a mean total energy for particle of $-10.00\mathscr{E}_0$.}
    \label{fig:419}
  \end{center}
\end{figure*}

\subsection{Validation of the trapped-electron detection algorithm: consistency within the simulation and benchmark against atomic-kinetic calculations}
\label{sec:ionization}

In order to validate the electron-trapping detection algorithm that we have established, we aim to compare its predictions. To this end, we will use a two-fold approach, checking the self-consistency of the simulations and benchmarking the obtained results against real atomic-kinetic calculations, to establish their validity for calculating Stark-broadened spectra.

To check the self-consistency within the simulation, we will compare the predictions from the electron-trapping detection algorithm with a different independent method, consistent with the \FMD framework. In this regard, we note that the new algorithm presented in this work can additionally be used to estimate the number of electrons that are trapped by an ion. As such, using this algorithm, one can easily determine the fractional population of different ion charge states in the plasma at given conditions and therefore obtain the corresponding ionization balance. 
It should be noted that this is not the main purpose of the proposed algorithm, which is intended to select meaningful field sequences for line broadening calculations. While it is possible to use it to determine the mean charge in some simple configurations (e.g. $Z=2$ ions), it is not expected to be an accurate method for determining this metric in higher charge systems.

Nevertheless, the plasma ionization balance can also be obtained from the analysis of the ion potential energy distribution. In a \FMD simulation with a given set of conditions, one can easily obtain the potential energy distribution of the plasma particles at any given time. By averaging over a sufficient number of instants and a sufficient number of simulation runs starting with different initial conditions ($\sim 10$), it is possible to obtain a good statistical representation of this distribution. 

\begin{figure}
    \centering
    \includegraphics[width=\columnwidth]{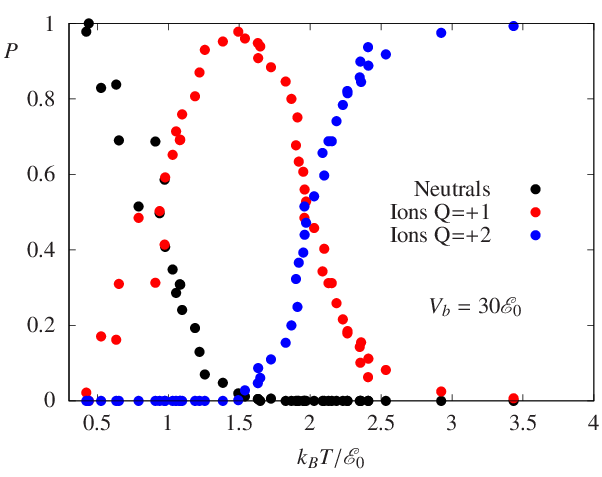}
	\caption{Fractional population of the different ion species as a function of the temperature for a plasma of ions with charge 2, and $V_b = 30\mathscr{E}_0$. Each dot corresponds to a set of simulations with the same temperature at equilibrium.}
    \label{fig:421}
\end{figure}

Using this method, we calculated the fractional population distribution for a collection of simulation runs spanning over a wide range of plasma temperatures for the case of $Z=2$ and $V_b=30\mathscr{E}_0$. The results are shown in Figure \ref{fig:421}, where the the different colors correspond to the population fraction of the different ion species as a function of temperature. We note that each dot in fact corresponds to a group of different simulations associated to the same temperature value, i.e. for a given temperature, a number of simulations starting from different initial conditions were launched and then let to thermalize following the method described in our previous work \citep{gigosos2018}. As expected, at low temperatures all ions have two bound electrons and, as the temperature increases, some electrons become free, leading to an increase in the population of ions with only one bound electron. The number of ions with $Q=+1$ keeps growing until almost no neutral ions remain. If the temperature is increased further, ions can become doubly ionized ($Q=+2$), until there are no more bound electron-ion pairs.
To illustrate how the ionization balance can be obtained from the potential energy distribution, let us first assume the simple case of hydrogen ions ($Z=1$). If the plasma temperature is low enough so that the distribution of kinetic energies is relatively narrow, the distribution of potential energies presents two main lobes. One of them centered around zero (free particles), and one of them near the potential well of energy $V_b$ (ions with a bound electron), as shown in figure 14 in \citet{gigosos2018}. In the same manner, we could expect that, for ions with $Z=2$, the distribution of potential energies can have up to three lobes: that corresponding to free particles, around zero potential energy; the one accounting for ions with one bound electron, with potential energies similar to $V_b$; and the lobe corresponding to ions with two trapped electrons, which will have potential energies similar to $2V_b$ (and equivalently for higher $Z$).

If these lobes are sufficiently separated, it is possible to integrate the energy distribution and associate the area under each lobe to the population of the different species. This is a robust method that does not require the definition of ad-hoc criteria to label an electron as bound or free. On the contrary, due to its own statistical nature, this method is not useful for selection of electric field sequences since the time-resolved information is lost. Hence, the comparison of the ionization balance obtained from the trapped-electron detection algorithm against the one from the analysis of the potential energy distribution provides a way to validate the former, adding confidence on the selected electric field sequences for line shape calculations.  

This is illustrated in Figure \ref{fig:418}, in which we show the distribution of potential energies (left) and its integral (right) for an equilibrated plasma of $Z=2$. The corresponding potential well is $V_b=30\mathscr{E}_0$, and the mean total energy per particle is $-2.12\mathscr{E}_0$. In this case, the energy per particle is relatively high compared with the potential well, and therefore only two charge species are present in the plasma: free ions and those with one bound electron. The distribution of potential energies therefore only presents two lobes, centered around zero and $V_i$. By looking at the integral of the energy distribution (plot on the right), the ratio of ions with one bound electron ($I_1$) to free particles ($I_0$) can be easily obtained.

A similar example is shown in Figure \ref{fig:419} for the case when the plasma mean energy is $-10\mathscr{E}_0$. The rest of the plasma parameters are identical as those in Fig. \ref{fig:418}. In this case, the plasma energy is too low, and the plasma is composed of only ions with one ($I_1)$ or two bound electrons ($I_2$), with almost no free ions.

\begin{figure}
    \centering
    \includegraphics[width=\columnwidth]{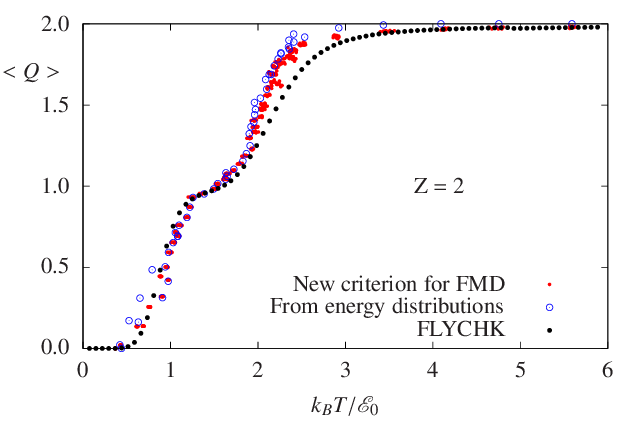}
    \caption{Mean ion charge obtained by applying the new criterion for selecting electric field sequences, compared to the values obtained by integrating the potential energy distributions. It can be seen that the results are in very good agreement.}
    \label{fig:cargaMedia}
\end{figure}

In Figure \ref{fig:421} we show the fractional population of each charge species for a $Z=2$ with $V_b = 30\mathscr{E}_0$ obtained from a batch of simulations with different temperatures.
Once the ion population distribution is obtained, one can easily obtain the average ionization, which permits the comparison with the results from the application of our trapped-electron detection algorithm. We show this in Figure \ref{fig:cargaMedia} where the results from the two methods are shown for the same temperature range. Red points correspond to the values obtained with the electron capture detection algorithm, whereas blue circles correspond to those obtained from the analysis of the potential energy distribution. The two methods show good agreement, which indicates that the results produced by our proposed algorithm are reasonable and self-consistent within the simulation.

Additionally, as mentioned above, when the temperature becomes even one fourth of the ionization potential ($\sim 5\mathscr{E}_0$), it can be seen that most atoms have $Z=2$, and therefore do not have captured electrons orbiting them so that recombination effects become negligible, which validates our estimation of $\tau_{bound}$.

To further benchmark our method, we compared our obtained values with the results from atomic kinetics calculations run with the collisional-radiative code FLYCHK \cite{chung2005} in local thermodynamic equilibrium conditions. These calculations use the quantum Saha equation, which uses Planck's constant to determine the size of the unit cell in phase space.
Naturally, this constant does neither appear in the classical simulations, nor in the equilibrium equations derived in Reference \cite{gigosos2018}. 
It was found that the values for the recombination rates obtained from the collisional-radiative calculations were in agreement with those obtained in the classical simulations within a factor of $\sim 2$, which is common even among different atomic kinetic codes.
The ionization values obtained with FLYCHK are indicated with the black dots in Figure \ref{fig:cargaMedia}, showing good agreement with the values obtained from the classical simulation. As mentioned above, this agreement validates the choice of $V_b$ given in Equation \ref{eqn:V0}.

\section{Conclusions}
\label{sec:Conclusions}

In this study, we have introduced a robust criterion for tracking and identifying trapped electrons in \textit{Full Molecular Dynamics} (FMD) computer simulations, focusing on ions with charges $Z>1$. Our presented criterion effectively removes the unphysical field sequences produced by orbiting electrons, allowing the selection of appropriate time-sequences of the electric field felt by a radiator for Stark broadening calculations. Additionally, this method offers a pathway to determine the plasma ionization balance under specific conditions.  

To validate our detection algorithm, we conducted extensive comparisons across a wide range of plasma temperatures. Specifically, we compared the average ionization values obtained through our criterion with those derived from a well-established method based on the analysis of ion potential energy distribution. While our ionization balance calculations do not claim to precisely solve plasma atomic kinetics, our \FMD simulations accurately capture the plasma ionization balance within the framework of the classical ionization-recombination mechanism.
Future work includes studying the sensitivity of these results with the choice of $\tau_{bound}$.

In addition, we compared our results with those obtained with the collisional-radiative code FLYCHK, finding remarkably good agreement between the classical simulation and the atomic-kinetics calculations both for the ionization balance and the lifetime of the charge states.

Our simulations, though comprehensive, do not encompass dissipative radiative processes. Consequently, the classical recombination mechanism incorporated in our simulations cannot fully account for radiative recombination; rather, it represents a combined effect arising from electron capture and likely three-body recombination. Nevertheless, the detection algorithm discussed in this study serves as a valuable tool for tracking the ionization-recombination mechanism within full molecular dynamics simulations involving ions with $Z>1$. This inclusion is pivotal, especially when considering the additional recombination width on spectral line shapes.

Importantly, without a methodology akin to the one presented in this study, such effects would be entirely overlooked, leading to inaccurate calculations of line profiles, particularly in strongly-coupled plasmas. By incorporating our detection algorithm, researchers can achieve more precise and reliable simulations, essential for advancing our understanding of spectral line broadening in complex plasma environments.

It is worth mentioning that the criterion presented in this work is applicable when the potential regularization radius $\rmin$ is smaller than the average distance between particles $r_e$. While this is most often the case, for heavily coupled systems such as solid-density plasmas \cite{vinko2012}, care must be taken before enforcing such criterion. Future work on this topic will include the study of \FMD simulations techniques for this type of crystalline systems.

\section*{Acknowledgements}

This work has been supported by Research Grants No. ENE2015-67581-R/FTN (MINECO/FEDER–UE) from the Spanish Ministry of Economy and Competitiveness and PID2022-137632OB-I00 from the Spanish Ministry of Science and Innovation.

This work has also been carried out within the framework of the EUROfusion consortium, funded by the European Union via the Euratom Research and Training Program (Grant Agreement Nos. 633053 and 101052200—EUROfusion). The views and opinions expressed are, however, those of the author(s) only and do not necessarily reflect those of the European Union or the European Commission. Neither the European Union nor the European Commission can be held responsible for them. The involved teams have operated within the framework of the Enabling Research Projects: Grant number AWP21-ENR-IFE.01.CEA. and AWP24-ENR-IFE.02.CEA-01.

\bibliography{biblio}

\end{document}